# Extraction of diffusion coefficients from the study of Rb release in different carbon catchers

Julien Guillot[a*], Brigitte Roussière[a], Pascal Jardin[b], Emeline Charon[c], Ugo Forestier-Colleoni[c], Romain Lafourcade[c], Martine Mayne-L'Hermite[c], Elie Borg[a], Vincent Bosquet[b], François Brisset[d], Wenling Dong[a], Stéphane Jourdain[a], Matthieu Lebois[ae], Damien Thisse[a]

[a] Université Paris Saclay, CNRS/IN2P3, IJCLab, 91405 Orsay, France

[b] GANIL (Grand Accélérateur National d'Ions Lourds), 14076 Caen Cedex 5, France

[c] Université Paris-Saclay, CEA, CNRS, NIMBE, 91191, Gif sur Yvette Cedex, France

[d] Université Paris Saclay, CNRS, ICMMO, 91405 Orsay, France

[e] Institut Universitaire de France, 1 rue Descartes, 75005 Paris, France

*Corresponding author: julien.guillot@ijclab.in2p3.fr, tel: +33169157248

______________________________________________________________________________________

______________________________________________________________________________________

**Abstract**:

New Target-Ion Source Systems combining a target and a catcher material are developed in the radioactive beam community, in particular at GANIL, in order to maximise the yield of very short lived atoms by minimizing the atom-to-ion transformation time. The aim of this study is to characterize the release properties of $^{81}$Rb collected on two graphite catchers and two carbon nanotube catchers. The release fractions were measured at various catcher-heating temperatures and then compared to the analytical expressions relevant to each catcher. This comparison led to the extraction of the pre-exponential factor ($D_0$) and the activation energy ($E_{act}$) involved in the diffusion coefficient of Rb for three carbon microstructures. All these data allowed to define an ideal catcher which could be made of aligned carbon nanotubes of small diameter and oriented in order to collect all the $^{81}$Rb atoms produced by the target but also to release them efficiently.

______________________________________________________________________________________

1. Introduction

The study of exotic nuclei is fundamental in nuclear physics to improve our knowledge and understanding of nuclear systems. Pushing away the limits of our knowledge requires to produce always more exotic nuclei and thus of shorter half-life. The ongoing research and development program at the SPIRAL1 facility (Système de Production d'Ions Radioactifs Accélérés en Ligne) at GANIL (Grand Accélérateur National d'Ions Lourds) in Caen aims to provide new and intense beams of exotic nuclei [1] using the Isotope Separator On Line (ISOL) method. This method consists in producing radioactive nuclei in a target and stopping them in a matrix, which can be the target material itself or another material, called catcher if separated from the target. Once stopped, the nuclei are neutralized and become atoms. The stopping material is generally maintained at high temperature to accelerate the atomic diffusion out of the material. Once released, atoms effuse to be ionized by an ion source and accelerated by an electric field to form a radioactive ion beam. To maximize the ion intensities, losses during all the atom-to-ion transformation (AIT) process must be minimized. For this, AIT time within the production system must be as short as possible compared to the collected nuclei decay time. This type of Target-Ion Source Systems (TISS) using catcher technology has been studied and developed at IGISOL in Jyväskylä [2] and FRIB at Michigan State University [3].

The new TISS developed at GANIL, called TULIP (Target Ion Source for Short-Lived Isotope Production), combines a target and a catcher material, and aims at minimizing the AIT time by optimizing each step of the AIT process (see ref. [4] for more details on the approach). The first objective of the TULIP project is to produce short-lived alkali isotopes of $^{74}$Rb ($T_{1/2}$ = 65 ms). They are produced by fusion-evaporation reactions induced by collision of stable beam at an energy close to the Coulomb barrier with a solid target. The latter is thin enough to allow produced nuclei to pass through. They are then implanted in the first tens of micrometers of the catcher material, depending on its density, and are neutralized. Then Rb atoms released out of the catcher effuse in the TISS cavity and are ionized by surface ionization to form a radioactive ion beam.

Graphite sheet of 200 μm was initially considered as a good candidate for the catcher for technical reasons of resistivity, heat resistance, ease of use, which facilitate the R&D of the TISS, and also for physico-chemical reasons related to its porosity and to the short sticking time of Rb on graphite [5] at the working temperature required for the TISS (~1400°C). Despite mechanical features less adapted to its use in the TULIP TISS design, rigid graphite of 1 μm grains used to design graphite targets of the current FEBIAD and Nanogan TISS at GANIL [6] was also selected for comparison. Two types of carbon nanotubes were also considered owing to their microstructure, i.e. dense tubes separated by large straight regions free of matter, which should favour a rapid effusion of the atoms out of the structure.

The objective of the present work was to determine which catcher material among the four selected offered the fastest release of Rb. The experiment took place at the Tandem accelerator on the ALTO (Accélérateur Linéaire et Tandem à Orsay) platform of Irène Joliot-Curie Laboratory (IJCLab). Our measurement method of the released fractions is an off-line procedure that can only be applied with relatively long-lived isotopes. Thus $^{81}$Rb ($T_{1/2}$ = 4.57 h) has been used in the present study. From the results of release versus temperature and with certain assumptions about the microstructure of the material and the implantation depth, pre-exponential factor ($D_0$) and activation energy ($E_{act}$) involved in the diffusion coefficient of Rb are extracted. As diffusion properties depend in a negligible way on the isotope of the element, diffusion parameters deduced for $^{81}$Rb have been used to evaluate release time and efficiency for $^{74}$Rb.

2. Presentation of the catcher materials

The release properties of four carbon samples, potential candidates for the TULIP catcher, were investigated. The catchers studied are two graphite samples that will be hereafter referred to as Papyex®, POCO and two samples of carbon nanotubes aligned either vertically (CNT$_v$) or horizontally (CNT$_h$) with respect to the substrate, placed perpendicularly to the ion beam. Figure 1 shows the scanning electron microscopy (SEM) images of the different samples.

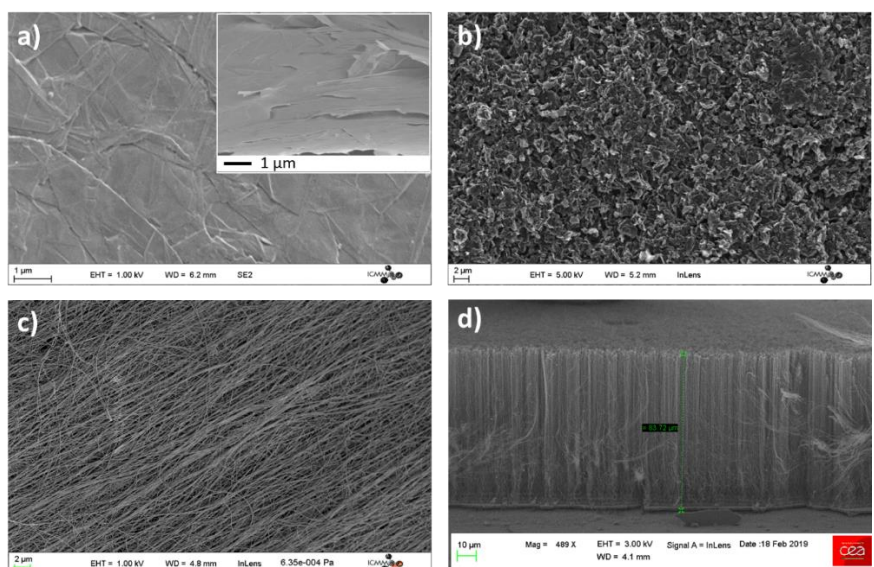

Figure 1 : SEM images of the catchers used in this experiment
a) Papyex® b) POCO c) Carbon nanotubes aligned horizontally (CNT$_h$)
d ) Carbon nanotube aligned vertically (CNT$_v$).

All these carbons have different physicochemical characteristics and were characterized when possible by Brunauer, Emmett and Teller (BET) method to determine the specific surface area (SSA) and by helium pycnometry to determine the density and percentage of open and closed porosity.

The Papyex® samples are from a 200 µm thick flexible graphite sheet with anisotropic grain orientation sold by Mersen industry under the name Papyex® (figure 1a). According to the figure 1a, the Papyex® sample is made of very thin sheets of micrometric size in surface and nanometric in thickness. We determined that the material constituting this sample has an apparent density of 1.15 g/cm$^3$ and has 44 % open porosity and 5 % closed porosity. This sample develops a specific surface area of 22.66 m$^2$/g.

The graphite samples POCO-ZXF-5Q1 (Entegris, USA) below simply referred to as POCO samples are isomoulded graphite carbon pieces (figure 1b). The samples are 440 µm thick with an apparent density of 1.82 g/cm$^3$, having respectively 14 % and 5 % of open and closed porosity. This sample has a SSA of 0.72 m$^2$/g, which corresponds to spherical grains of 3.7 µm in diameter.

The Papyex® and POCO samples were obtained by cutting from the sheets 2 cm by 1.5 cm rectangles suitable for the sample holder.

The CNT$_h$ samples are obtained from sheets composed of carbon nanotubes horizontally aligned (i.e. parallel to the faces of the sheet) and sold by the Merck company (901082-1EA, Merck) as displayed in figure 1c. These sheets of carbon nanotubes were synthesized according to the protocol developed by Inoue *et al.* [7]. The carbon nanotubes have a SSA of 97.25 m$^2$/g, an average diameter of 46 nm and a length of 2 mm forming a sheet of about 3 µm thick with an average apparent density of 0.398 g/cm$^3$. The CNT$_h$ samples correspond to a stack of 4 sheets of horizontally aligned carbon nanotubes representing a mass per unit area of 0.456 mg/cm$^2$ ± 0.024 mg/cm$^2$, i.e. a total thickness of about 11.5 µm, deposited on a sheet of Papyex® pre-cut to form rectangles of 2 cm by 1.5 cm.

The CNT$_v$ samples were synthesized and characterised by the NIMBE-LEDNA research team (CEA Paris-Saclay, France). These carbon nanotubes have been synthesized by aerosol-assisted catalytic chemical vapour deposition on a 1.5 cm by 1.5 cm quartz substrate with a thickness of 1 mm [8]–[10]. The carbon nanotubes have a length of 90 µm for an external diameter of 10 nm (figure 1d). The apparent density of this material is 0.3 g/cm$^3$ and the CNT number density is less than 10$^{11}$ CNT/cm$^2$. Taking into account the microstructure of the catcher (vertically aligned CNTs which do not form a closed cavity), the part of the catcher not occupied by the CNTs (≥ 92 % of the total surface) can be associated to an open porosity. SSA measurements were not carried out for this catcher because it implied to separate the CNTs from the quartz support, which was not possible due to the small number of samples available.

The physical properties of the four carbon catchers are summarized in Table 1.

Table 1: Physical properties of the four carbon catchers; in the third column, L, l and d indicate the length, width and thickness of the catcher

| Catcher | Substrate | Catcher dimension L (cm) × l (cm) × d (µm) | Density (g/cm$^3$) | Porosities (%) open | Porosities (%) close | SSA (m²/g) | CNT properties Diameter | CNT properties Length | CNT properties Number |
|---|---|---|---|---|---|---|---|---|---|
| Papyex® | - | 2 × 1.5 × 200 | 1.15 | 44 | 5 | 22.66 | - | - | - |
| POCO | - | 2 × 1.5 × 440 | 1.82 | 14 | 5 | 0.72 | - | - | - |
| CNT$_h$ | Papyex® | 2 × 1.5 × 11.5 | 0.398 | - | - | 97.25 | 46 nm | 2 mm | - |
| CNT$_v$ | Quartz | 1.5 × 1.5 × 90 | 0.3 | 92* | - | - | 10 nm | 90 µm | < 10$^{11}$ cm$^{-2}$ |

* part of the catcher not occupied by the CNTs

3. Measurement of released fractions

The method used to measure the released fractions (*RF*) is adapted from the protocol developed for the research and development on ISOL targets at IJCLab [11], [12]. A copper target ($^{65}$Cu) 2.72 µm ± 0.29 µm thick was irradiated for 30 minutes with a $^{19}$F$^{6+}$ beam of 62.5 MeV energy and 20 nAe intensity. $^{81g}$Rb and $^{81m}$Rb are produced by fusion evaporation and are stopped in a catcher sample located 6 mm behind the target. The target and the catcher sample are placed in a sample holder which allows the adjustment of the $^{19}$F beam before bombarding the target. To do so, in addition to the target, a scintillator with a 9.5 mm hole is place in the same sample holder. The latter can be moved perpendicularly to the beam axis. When the scintillator faces the beam, the position, size and shape of the beam are adjusted and when the hole is on the beam axis, the ion beam intensity is measured. The beam has the shape of an ellipse with a 2.30 mm major and a 1.75 mm minor axes. The rate of the isotopes implanted in the catcher was obtained by γ spectrometry measurements: the transitions characteristic of the $^{81}$Sr, $^{81m}$Rb and $^{81}$Rb disintegrations are clearly observed in the spectra recorded with a waiting time between the end of the irradiation and the start of the counting below 45 m and allow us to determine that $(2.2 \pm 0.2) \times 10^4$ $^{81}$Sr, $(6.6 \pm 0.9) \times 10^4$ $^{81m}$Rb and $(2.6 \pm 0.4) \times 10^4$ $^{81}$Rb were implanted per second in the catcher. $^{81}$Sr ($T_{1/2}$ = 22.3 m) decaying into $^{81}$Rb, it was more advisable to wait for the complete decay of this element before the measurements, *i. e.* about 2h30 (corresponding to ~ 7 times the $^{81}$Sr half-life) of waiting after each irradiation. In this way, no $^{81}$Rb is created during the heating of the catcher and the release fraction determination is fair. After this waiting time, a first counting of the gamma catcher activity is performed at room temperature for 30 minutes using a germanium detector. At the end of the counting, the catcher is placed under vacuum ($10^{-6}$ mbar) in an oven previously calibrated between 250 °C and 1100 °C, temperature range in which $^{81}$Rb released fractions were estimated to be less than 100 %. The temperature of the oven was calibrated using thermocouples and different metals (Sn, Pb, Zn, Al, Cu) heated up to their melting points. The heating device was controlled by a power supply controller, which allowed a repeatability of the heating cycles, in time and power, and of the temperature measurement with ± 20 °C accuracy. After this 30-minute heating step, the catcher residual activity was determined by performing a second gamma spectrometry measurement for 30 minutes.

For each catcher the sample number available, the measurement number performed and the temperature range investigated are presented in Table 2.

Table 2: For each catcher, number of samples and of measurement points and temperature range studied

| Catcher | Carbon source | Number of samples | Number of measurements | Temperature range |
|---|---|---|---|---|
| Papyex® | Graphite | 10 | 10 | 250 °C – 965 °C |
| POCO | Graphite | 5 | 5 | 540 °C – 965 °C |
| CNT$_v$ | Carbon nanotube | 3 | 7 | 310 °C – 940 °C |
| CNT$_h$ | Carbon nanotube | 4 | 3 | 660 °C – 860 °C |

Figure 2 shows for the POCO catcher the γ spectra obtained before and after heating at different temperatures.

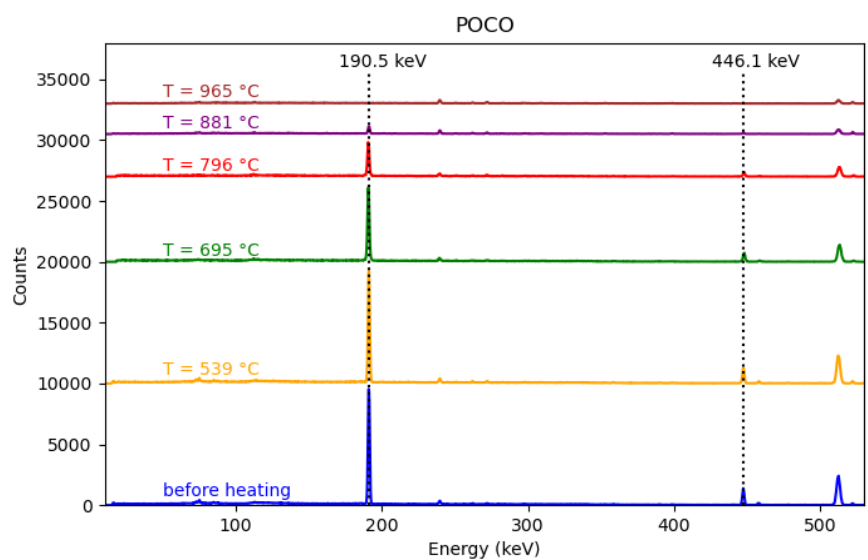

Figure 2: Spectra obtained by γ spectrometry before and after heating for different temperatures for POCO samples. An arbitrary constant was added to the heated spectra in order to obtain a more understandable figure.

This figure illustrates the influence of temperature on the intensities measured for the 190.5 and 446.1 keV γ transitions resulting from the $^{81g}$Rb decay.

The released fraction (*RF*) was obtained from Eq. (1):

$$RF(E_\gamma) = 100 \times \left(1 - \frac{I_{heated}(E_\gamma)}{I_{unheated}(E_\gamma) \times e^{-\lambda.t_w}}\right) \quad (1)$$

with $I_{heated}(E_\gamma)$, the intensity of the γ ray of $E_\gamma$ energy after heating, $I_{unheated}(E_\gamma)$, the intensity of the γ ray of $E_\gamma$ energy before heating, $\lambda$, decay constant and $t_w$, waiting time between the first and second counting.

Three RF values were obtained from the γ transitions signing the $^{81g}$Rb decay and observed in the γ spectra recorded (190.5, 446.1 and 456.7 keV). The adopted released fraction is the average of these three values weighted by the transition intensities.

4. Results

As shown in table 2, we did not have the same number of samples for each catcher and therefore we could not make the same number of measurements for all the catchers. We had 10 samples of Papyex®, we could make 10 measurements in a wide range of temperature between 250 °C and 965 °C and thus define the temperature range in which graphite releases Rb. For the POCO catcher, 5 measurements were done in a narrower temperature range between 540 °C and 965 °C. For the $CNT_v$ catcher, we had only 3 samples. In order to perform 7 measurements between 310 °C and 940 °C, each sample was irradiated twice with a 48 h delay between the irradiations so that the catcher no longer contained $^{81}$Rb atoms ($T_{1/2}$ = 4.5 h). The incident beam was slightly shifted between the two irradiations to avoid $^{81}$Rb being implanted in the same place in the catcher and to limit the possible damage caused by the particles entering the catcher. The sample heated to 310 °C did not release any $^{81}$Rb, it was used to carry out the measurement at 590 °C without new implantation. The 4 $CNT_h$ samples, made of horizontally-aligned-CNT sheets deposited on a support which is a Papyex®-sheet backing, can be used only once. After the irradiation of the first $CNT_h$ sample, we carried out a γ counting to determine the total amount of $^{81}$Rb collected in the catcher. The $CNT_h$ sheets were then separated from the Papyex® backing and the activity emitted by each of two pieces was evaluated by separate gamma spectrometry measurements. We so have determined that 57 % of the $^{81}$Rb atoms were stopped in the Papyex® backing and 43 % in the $CNT_h$ sheets. For this catcher, we are interested only by the release from the carbon nanotubes, thus a slightly different protocol has been applied to extract the relevant released fraction. For the three remaining samples, after the irradiation, the first γ counting and the heating, the $CNT_h$ sheets are separated from their Papyex® backing and the second γ counting is performed on the $CNT_h$ sheets alone and on the Papyex® backing separately. The released fractions are obtained by taking for $I_{unheated}$ in Eq. 1 only 43 % and 57 % of the intensity measured during the first counting, respectively. Figure 3 displays all the released fractions measured for the catchers studied during this experiment.

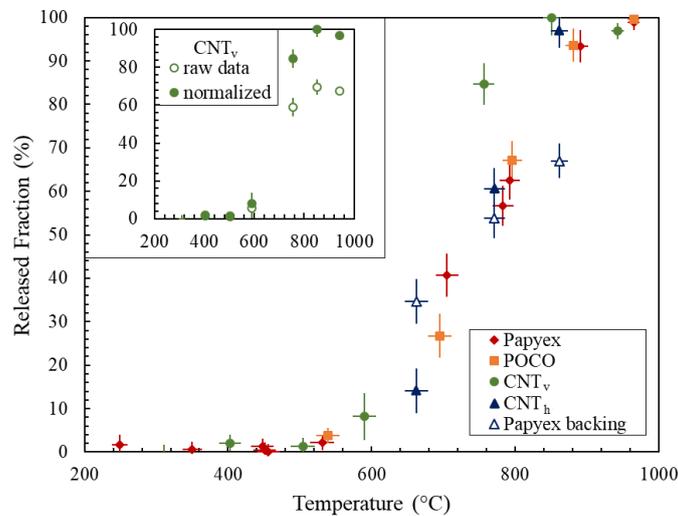

Figure 3: Released fractions measured for the catchers studied in this experiment. For the $CNT_v$, the raw and normalized data (see text) are presented in the insert.

Around 950 °C, all the $^{81}$Rb atoms stopped in the Papyex® POCO and CNT$_h$ catchers are released, but for the CNT$_v$ catcher, the released fractions measured do not exceed 69 % (see insert Fig 3). By similarity with the other catchers, we assume that at 943 °C all the $^{81}$Rb stopped in the CNT were also released. This implies that the activity still present in the catcher comes from the fraction of $^{81}$Rb (31 %) stopped in the quartz and not released at this temperature. As we are only interested by the release from the carbon materials, the released fractions have been normalized to 100 %. Surprisingly, compared to what is observed on the Papyex® catcher, the fractions released from the Papyex® backing are higher at low temperatures and lower at high temperatures. This will be discussed further below.

Taking into account the error bars, it results from Figure 3 that all catchers do not release $^{81}$Rb below 551 °C but release almost totally this element above 943 °C. In the temperature range where the release takes place, two categories of catcher seem to emerge. The first group includes the Papyex® and POCO catchers that exhibit very similar released fractions increasing almost linearly in the 590-891 °C temperature range. The second one consists of the two carbon nanotube based catchers, CNT$_v$ and CNT$_h$, that, compared to the Papyex® and POCO catchers, seem to be better above 771 °C. As it stands and given the low numbers of measurement points and the temperature uncertainty, it seems difficult to identify any impact of the CNT orientation on the release properties of the catcher.

5. Determination of the release time and efficiency

In order to compare the release properties of the studied catchers, release efficiencies are more convenient than released fractions. Both depend on the diffusion coefficient and can be calculated in the frame of models describing diffusion in specific media: in a sheet, a cylinder, a sphere or a fibre [12]–[14]. The choice of the medium in which the diffusion takes place will be made according to the structural properties of the different catchers. In the case of diffusion in a sheet, the profile and depth of implantation must also be known.

Profile and implantation depth were calculated in the frame of the LISE++ software (version 15.13.7) [15]. Then the relevant analytical expressions linking the diffusion coefficient to the released fractions and to the release efficiency for a radioactive isotope are indicated. The method used to extract the activation energy, $E_{act}$, and the pre-exponential factor, $D_0$, involved in the diffusion coefficient is described. Finally, the release properties of the catchers are extrapolated and compared at a temperature of 1400 °C, which is the temperature aimed for on-line experiments.

5.1. Implantation depth of $^{81}$Rb in the different catchers

The calculations of the implantation require as input parameters the characteristics of the target (nature, thickness), those of the primary beam (energy) and of the catcher (nature, density, thickness). The results of the implantation depth simulations are shown in Figure 4.

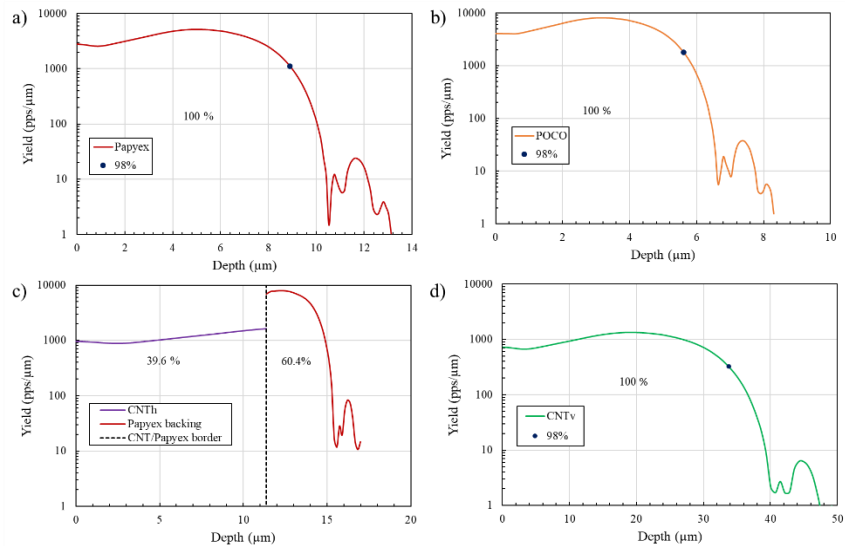

Figure 4: LISE ++ simulations of the Rb implantation depth in catchers with different densities
a) Papyex® b) POCO c) CNT$_h$ d) CNT$_v$

The LISE++ simulations indicate that 57 % of the reaction products exit the target. For the graphite catchers, 100 % of the $^{81}$Rb leaving the target are stopped and 98 % of them are implanted in the first 8.9 or 5.6 µm of the Papyex® or POCO catcher, respectively. For the CNT$_h$ catcher, 40 % of the $^{81}$Rb exiting the target are implanted in the carbon layer of 11.5 µm thick and 0.398 g/cm$^3$ density and the remaining 60 % in the Papyex® backing. The simulation results are in good agreement with the experimental measurements: 43 % in the CNT$_h$ layers and 57 % in the Papyex® sheet. For the CNT$_v$ catcher, LISE++ indicates that 100 % of the $^{81}$Rb are implanted in the CNT$_v$ catcher and 0 % in its quartz support, which is not in agreement with the experimental data (69 %, see section 4). However, the LISE++ code assumes that the materials have a uniformly distributed density, which is not the case for the CNT$_v$ catcher that is architecturally designed with the carbon nanotubes aligned perpendicularly to their support. The $^{81}$Rb beam is emitted in a truncated cone with a vertex angle of 74° inside which 34 % of the beam is collinear with the direction of the carbon nanotubes. Under these conditions, taking into account the diameter of the CNT$_v$ and of the Rb atom as well as the number of CNT$_v$ per unit area, only ~ 8 % of the $^{81}$Rb have a probability of encountering a carbon nanotube and implanting itself. Combining these two pieces of information, we can conclude that ~ 37 % of the Rb will be stopped in the quartz support. This latter estimate is in rather good agreement with the experimental values, i.e. the CNT$_v$ collects 69 % of the $^{81}$Rb and the quartz backing the remaining 31 %. The Rb trapping by quartz was already observed at ISOLDE in our temperature range [16]. Although the LISE++ simulations indicate that the $^{81}$Rb atoms are stopped in the catcher, all the available information (catcher architecture, Rb emission angle, quantity deposited in the quartz substrate) leads us to believe that Rb is also implanted in the region between 33.8 and 90 µm. As the catcher architecture is not taken into account by LISE++, the analysis was carried out for these two extreme implantation depths ($\delta_l$ = 33.8 µm and $\delta_2$ = 90 µm). Table 3 presents the results obtained by LISE++ and used for further analyses.

Table 3: Results of implantation of Rb obtained with the LISE++ simulation code. In the third column, the thickness calculated by LISE ++ and required to stop 98 % of the Rb produced is given. The fourth column gives the thickness of the catchers used. The fifth column gives the actual thickness of the material studied in which the Rb is stopped.

| Catcher | Density (g/cm$^3$) | Implantation depth (µm) | Catcher thickness " $d$ " (µm) | Real implantation depth " $\delta$ " (µm) |
|---|---|---|---|---|
| Papyex® | 1.15 | 8.9 | 200 | 8.9 |
| POCO | 1.82 | 5.6 | 440 | 5.6 |
| CNT$_h$ | 0.398 | 25.4 | 11.5 (CNT$_h$) + 200 (Papyex® backing) | 11.5 |
| CNT$_v$ | 0.3 | 33.8 | 90 (CNT$_v$) + 1000 (Quartz support) | 33.8 ($\delta_l$) 90 ($\delta_2$) |

5.2. Diffusion models used

The physicochemical (density, porosity) and structural characteristics (given by the SEM images) lead to the breakdown of the catchers into two groups. The first group includes catchers with a very compact sheet or grain structure, a "high" density (> 1 g/cm$^3$) and porosity including closed porosity. The second group comprises catchers showing a highly aligned fibre structure, a "low" density (< 0.4 g/cm$^3$) and with the additional characteristic that the tubes are separated from each other and therefore form neither open nor closed porosity.

These very marked differences between the two classes suggest that the relative weights of diffusion and effusion processes will not be the same in the two cases. The released fractions measured include all the phenomena occurring in the catchers, the diffusion in the material (carbon) and the effusion in the porosity. But, whatever the microstructure considered (sheet, fibre, cylinder, sphere) and the propagation process (diffusion in material or effusion between the micro-structure of material), the released fractions are analysed using a formalism generally used for diffusion. If the microstructure of the carbons that make up the catchers is not taken into account, the catchers can be analysed globally and assimilated to a sheet of the catcher dimensions. The model of diffusion in a sheet proposed by Fujioka and Arai [13] cannot be used here because the authors assume that, at t = 0, the atoms that will diffuse are uniformly distributed in the sheet and can exit by the two faces of the sheet. For the Papyex® and POCO catchers, the LISE++ calculations showed that the Rb atoms were stopped and distributed in the first µm of the catcher as shown in Figure 4. In the case of CNT catchers, the experiment has shown that Rb are implanted not only along the CNT thickness but also in the Papyex® or quartz support. In the case of the $CNT_v$, the quartz has not released the implanted Rb and forms a barrier for the Rb implanted in the CNT which can only exit through the free face. In the case of the $CNT_h$ catcher, the Papyex® support has released the implanted Rb but the release profile is very different from that of the Papyex® catcher (figure 3) indicating that there is an exchange between the Papyex® support and the $CNT_h$. A sheet model with a homogeneous implantation and an exit by only one face is not sufficient to describe the process involved in the $CNT_h$ catcher. Therefore, this catcher will not be considered in the rest of the study.

We have established the relation allowing us to calculate the released fraction and the release efficiency of a radioactive isotope in both cases: diffusion in a sheet with homogeneous implantation in surface and exit by one or two faces.

### 5.2.1. Relation between released fractions, release efficiencies and diffusion coefficients

The diffusion coefficient ($D$) is described by an Arrhenius-type equation and is written as follows (equation 2):

$$D = D_0 \, exp\left(-\frac{E_{act}}{kT}\right) \quad (2)$$

with $D_0$ the pre-exponential factor, $E_{act}$ the activation energy, $T$ the temperature and $k$ the Boltzmann constant.

### 5.2.2. Diffusion in a sheet with exit by one or two faces

The analytical expressions for the released fraction and the release efficiency of a radioactive isotope were established following the method of separation of variables described by Crank [17]. The boundary conditions were chosen to describe the experimental conditions.

In the case of an implantation near the surface and exit by two faces, boundary conditions were: at t = 0 the amount of atoms implanted is uniform between *0* and *δ* and for t > 0, the amount of atoms is zero at positions *0* and *d*. *d* is the target thickness and *δ* the implantation depth. This latter boundary condition means that the diffusing atom, having arrived at the foil surfaces, is released very quickly.

In the case of exit by one face, boundary conditions are: at *t = 0* the amount of atoms implanted is uniform between *0* and *δ* the implantation depth ($C(x) = C_0$ for $0 < x < δ$) and for *t > 0*, the amount of atoms is zero at position *0* indicating that the atoms exit through the face *x = 0* ($C(0) = 0$) and the condition to be satisfied at the impermeable boundary at the position *x = d* is $\frac{\partial C}{\partial x} = 0$, *d* being the thickness of the catcher.

The released fraction during a heating time $t_h$ is written (equation 3):

$$RF(t_h) = 1 - \frac{\sum_{m=1}^{\infty} \frac{1}{(2m-1)^2}\left(1-\cos\frac{(2m-1)\pi\delta}{\alpha d}\right)\exp\left(-\frac{(2m-1)^2\pi^2 D t_h}{(\alpha d)^2}\right)}{\sum_{m=1}^{\infty} \frac{1}{(2m-1)^2}\left(1-\cos\frac{(2m-1)\pi\delta}{\alpha d}\right)} \quad (3)$$

with $D$ the diffusion coefficient, $d$ the target thickness and $δ$ the implantation depth. α is a dimensionless coefficient equal to 1 in case of exit by two faces and equal to 2 in case of exit by one face.

The release efficiency ($\varepsilon_{RF}$) of a radioactive isotope is written (equation 4):

$$\varepsilon_{RF}(\lambda) = \frac{\sum_{m=1}^{\infty} \frac{\pi^2 D}{(2m-1)^2\pi^2 D + \lambda(\alpha d)^2}\left(1-\cos\frac{(2m-1)\pi\delta}{\alpha d}\right)}{\sum_{m=1}^{\infty} \frac{1}{(2m-1)^2}\left(1-\cos\frac{(2m-1)\pi\delta}{\alpha d}\right)} \quad (4)$$

with $\lambda$ the radioactive constant of the isotope considered.

The *d* and *δ* values used for the calculations are indicated in Table 3. Like in Fujioka and Arai's article [10], the foil dimensions other than thickness are assumed to be infinite. Indeed, given the dimensions of the catchers and the implantation profiles obtained with LISE++, almost all atoms will diffuse and exit through the faces perpendicular to the primary beam direction.

### 5.3. Determination of the constants $D_0$ and $E_{act}$

Knowing the characteristics of the sample studied (thickness of the sheet and depth of implantation) and the chosen diffusion model, the change in $RF$ as a function of $D$ is calculated. From this curve, a value $D_{exp}$ is assigned to each experimental value of $RF$. The parameters $E_{act}$ and $D_0$ are obtained from $D_{exp}$ values by performing a linear regression using the relation $Ln(D_{exp}) = Ln(D_0) - E_{act}/kT$. Finally, the experimental points $RF_{exp}$ and the calculated curve $RF_{calc}(E_{act}, D_0)$ are plotted as a function of temperature. As can be seen in Figure 3, for the three catchers, no release is observed below 550 °C and full release is obtained above 950 °C. It is just in this temperature range that the linear behaviour of $Ln(D_{exp})$ as a function of $1/T$ is observed. Therefore, fits were performed including only the experimental points verifying this linear behaviour.

It is worth noting that for a given value of $D$, the variables $D_0$ and $E_{act}$ are not independent (figure 5).

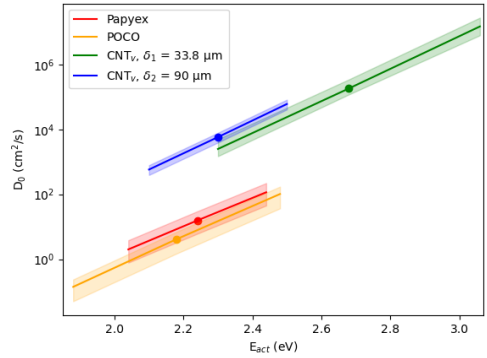

Figure 5: Variation of $D_0$ as a function of $E_{act}$ for Papyex®, POCO and $CNT_v$ catchers. For the $CNT_v$ catcher, the result obtained with the two analyses ($\delta_1 = 33.8$ μm and $\delta_2 = 90$ μm) are plotted. The dots represent the $E_{act}$ mean values with the associated $D_0$ taken in the rest of the study.

For each catcher various pairs of ($E_{act}$, $D_0$) give the same fit of the experimental RF values, for example:
- for POCO: (1.88 eV, 0.14 cm²/s), (2.18 eV, 4.1 cm²/s) and (2.48 eV, 100 cm²/s)
- for Papyex®: (2.04 eV, 2.01 cm²/s), (2.24 eV, 15.6 cm²/s) and (2.44 eV, 116 cm²/s)
- for $CNT_v$ ($\delta_1 = 33.8$ μm): (2.30 eV, 0.025×10⁵ cm²/s), (2.68 eV, 1.84×10⁵ cm²/s) and (3.06 eV, 155×10⁵ cm²/s)
- for $CNT_v$ ($\delta_2 = 90$ μm): (2.10 eV, 0.58×10³ cm²/s), (2.30 eV, 5.9×10³ cm²/s) and (2.50 eV, 61×10³ cm²/s)

It is worth noting that the variation of $D_0$ as a function of $E_{act}$ is very fast (fig. 5). For the $CNT_v$ catcher analyzed with $\delta_1 = 33.8$ μm, the $E_{act}$ range is higher than for the other catchers leading to a very large $D_0$ range, spreading over 3 orders of magnitude. The envelope drawn around the $D_0 = f(E_{act})$ curve shows the $D_0$ variation associated with a given $E_{act}$ value allowing to keep the calculated $RF$ values within the experimental error bars (represented on Figure 3). Table 4 shows, for each catcher, the diffusion model used, the $E_{act}$ mean value and the $D_0$ associated with its error bar.

Table 4: $D_0$ and $E_{act}$ values determined for each catcher within the temperature range of 550 °C to 950 °C

| Catcher | Density (g/cm³) | Model | Parameters | $D_0$ (cm²/s) | $E_{act}$ (eV) |
|---|---|---|---|---|---|
| POCO | 1.82 | Sheet with exit by two faces | $\delta = 5.6$ μm<br>$d = 200$ μm | $4.1 \pm 2.8$ | 2.18 |
| Papyex® | 1.15 | Sheet with exit by two faces | $\delta = 8.9$ μm<br>$d = 440$ μm | $15.6^{+14}_{-10}$ | 2.24 |
| $CNT_v$ | 0.3 | Sheet with exit by one face | $\delta_1 = 33.8$ μm<br>$d = 90$ μm | $(1.84^{+1.66}_{-0.64}) \times 10^5$ | 2.68 |
| | | | $\delta_2 = 90$ μm<br>$d = 90$ μm | $(5.9^{+2.1}_{-1.9}) \times 10^3$ | 2.30 |

Figure 6 shows for each catcher the comparison between the experimental released fractions (red dots) and the values calculated using the $D_0$ and $E_{act}$ parameters reported in Table 4 (blue dotted lines).

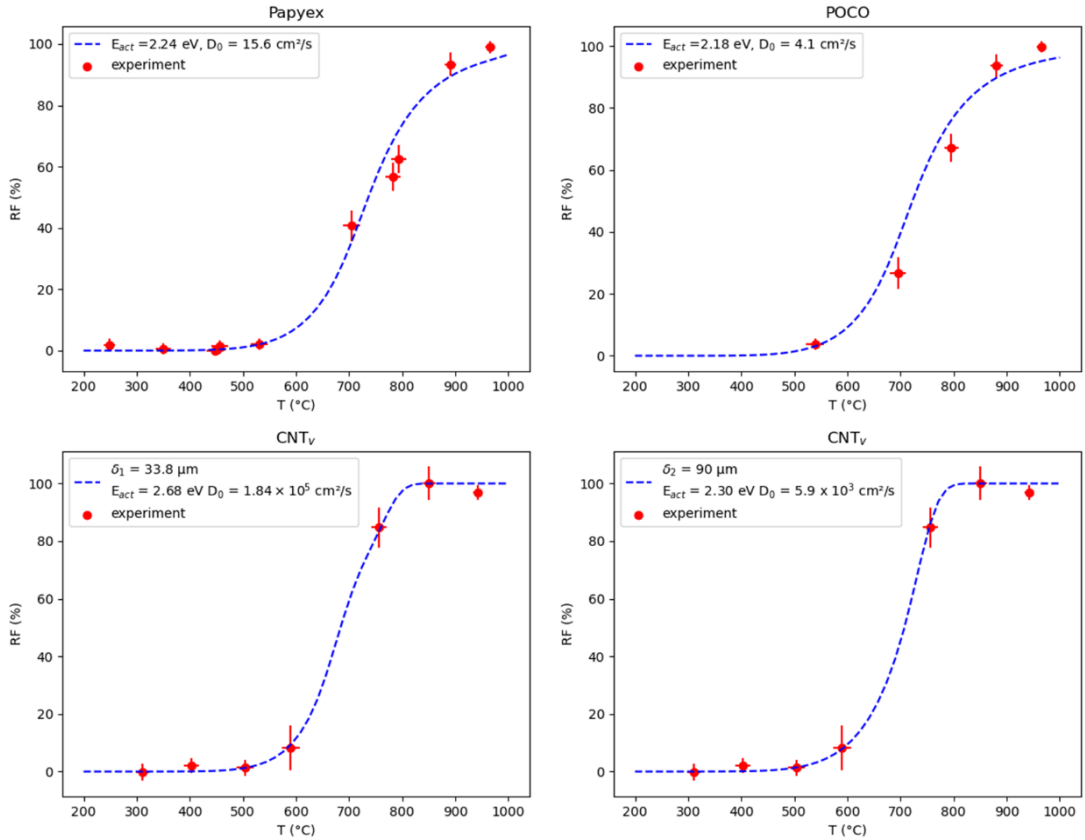

Figure 6: Comparison between the values of the experimental (red dots) and calculated (blue dotted lines) released fractions.

It results from table 4 and figure 5 that the $E_{act}$ values are close for the three catchers and are compatible within error bars. The lower the density of the catchers, the greater the $D_0$ values. These remarks suggest that the activation energy, mainly related to the barrier the atoms have to overcome in the material, does not change from one catcher to the other as all are based on the same atomic bond, and thus slightly depends on the density or on the space free of carbon matter. *A contrario*, the evolution of $D_0$ versus density suggests that $D_0$ is mainly related to the microstructure of the catchers free of carbon material (the density of the material associated with an open porosity ≥ 92 % for the $CNT_v$ and equal to 44 % and 14 % for Papyex® and POCO respectively).

6.  Estimates of the catcher release efficiencies

As the temperature of the TULIP-TISS is expected to be 1400 °C during on-line experiments, the diffusion coefficients of the different catchers have been extrapolated to this temperature assuming that the $D_0$ and $E_{act}$ values determined in the 550-950 °C range can still be used at 1400 °C. The diffusion coefficients, the time required for half of the stable nuclei present at t = 0 to be released ($t_{50}$) and the release efficiency of $^{74}$Rb ($\varepsilon_R(^{74}Rb)$) with half-life $T_{1/2}$ = 65 ms calculated at T = 1000 °C and extrapolated at T = 1400 °C are presented in Table 5.

Table 5: Diffusion coefficients (*D*), time to release 50 % atoms implanted (*t₅₀*) and $^{74}$Rb release efficiency ($\varepsilon_R(^{74}Rb)$) calculated at 1000 °C for the different catchers from the $E_{act}$ and $D_0$ values obtained during the analysis of the released fractions and extrapolated to a temperature of 1400 °C.

| Catcher | T (°C) | D (cm²/s) | $t_{50}$ (s) | $\varepsilon_R(^{74}Rb)$ |
|---|---|---|---|---|
| POCO $E_{act}$ = 2.18 eV $D_0$ = 4.1±2.8 cm²/s | 1000 | $(9.7 \pm 6.5) \times 10^{-9}$ | $7.5^{+13}_{-3}$ | $0.05 \pm 0.02$ |
| | 1400 | $(1.1 \pm 0.7) \times 10^{-6}$ | $0.07^{+0.12}_{-0.03}$ | $0.48^{+0.08}_{-0.15}$ |
| Papyex® $E_{act}$ = 2.24 eV $D_0$ = $15.6^{+14}_{-10}$ cm²/s | 1000 | $(2.1^{+1.9}_{-1.3}) \times 10^{-8}$ | $8.7^{+13.5}_{-4.1}$ | $0.05 \pm 0.02$ |
| | 1400 | $(2.8^{+2.6}_{-1.7}) \times 10^{-6}$ | $0.07^{+0.10}_{-0.03}$ | $0.47^{+0.10}_{-0.14}$ |
| CNT$_v$ (δ = 33.8 μm) $E_{act}$ = 2.68 eV $D_0$ = $(1.84^{+1.66}_{-0.64}) \times 10^5$ cm²/s | 1000 | $(4.5^{+4.1}_{-1.5}) \times 10^{-6}$ | $0.6 \pm 0.3$ | $0.19^{+0.07}_{-0.03}$ |
| | 1400 | $(1.6^{+1.4}_{-0.5}) \times 10^{-3}$ | $0.0017^{+0.001}_{-0.001}$ | $0.92 \pm 0.03$ |
| CNT$_v$ (δ = 90 μm) $E_{act}$ = 2.30 eV $D_0$ = $(5.9^{+2.1}_{-1.9}) \times 10^3$ cm²/s | 1000 | $(4.6^{+1.7}_{-1.5}) \times 10^{-6}$ | $3.5^{+1.6}_{-0.9}$ | $0.07 \pm 0.01$ |
| | 1400 | $(6.9^{+2.5}_{-2.2}) \times 10^{-4}$ | $0.023^{+0.011}_{-0.007}$ | $0.72^{+0.05}_{-0.08}$ |

For all catcher types, when the temperature increases from 1000 to 1400 °C, the diffusion coefficient increases by two orders of magnitude, $t_{50}$ decreases by two orders of magnitude and the $^{74}$Rb release efficiency increases by a factor of ~10. Although the diffusion coefficient is greater for Papyex® than for POCO, the $t_{50}$ and the release efficiencies are practically the same. In particular, the calculated release efficiencies at 1400 °C for $^{74}$Rb are in the order of 40-50 %.

This similarity may seem surprising at first sight. The SSA of Papyex® (22.66 m²/g) is greater than that of POCO (0.72 m²/g), reflecting a finer-grained microstructure, which should lead to a greater diffusion efficiency since the diffusion efficiency is inversely proportional to the grain size [13]. Moreover, the porosity of Papyex® (44 %) is high compared to that of POCO (14 %) and corresponds to a lower density (1.15 g/cm³ for Papyex® and 1.82 g/cm³ for POCO). The Rb atoms are therefore implanted more deeply in Papyex® than in POCO (see Figure 4), enlarging their effusion time out of Papyex®. As the effusion efficiency is inversely proportional to the effusion time [18], a greater effusion efficiency is expected in POCO than in Papyex®. The similarity between the calculated release efficiencies of $^{74}$Rb out of Papyex® and POCO materials thus probably results from a different time sharing between the diffusion and effusion processes, i.e. a shorter diffusion time but larger effusion time in Papyex® than in POCO.

The CNT$_v$ catcher has more pronounced differences in its structural features governing either the diffusion or the effusion process. In the first case, the size of the tubes is nanometric (10 nm in diameter) and in the second case the porosity is very high (≥ 92 %) and the density very low (0.3 g/cm³) which leads to an implantation over the whole thickness of the catcher. One can wonder how these properties will interact in the release process and whether one of the components (diffusion or effusion) will be predominant in the overall process. To answer this question, the analysis of the CNT$_v$ released fractions was performed using as model the diffusion in a cylinder [12], [14] with the same radius and length as the CNTs making up the carpet (*R* = 5 nm and *L* = 90 μm). Choosing this model means that the release of the atoms out of the catcher is only governed by the diffusion out of the cylinder material, and that the contribution of the effusion between the cylinders is negligible. In other words, the Rb atom is considered to leave the catcher when it leaves the tube. The figure 7 and the table 6 show the result of this analysis.

Table 6: $D_0$ and $E_{act}$ values determined for $CNT_v$ catcher within the temperature range of 550 °C to 950 °C in the frame of the cylinder model. Diffusion coefficients (*D*), time to release 50 % atoms implanted ($t_{50}$) and $^{74}$Rb release efficiency ($\varepsilon_R(^{74}Rb)$) calculated at 1000 °C and extrapolated to a temperature of 1400 °C.

| $E_{act}$ (eV) | $D_0$ (cm²/s) | T (°C) | D (cm²/s) | $t_{50}$ (s) | $\varepsilon_R(^{74}Rb)$ |
|---|---|---|---|---|---|
| 2.42 | $(2.7 \pm 2.0) \times 10^{-5}$ | 1000 | $(7.1^{+3.6}_{-2.2}) \times 10^{-15}$ | $2.2^{+1.3}_{-1.0}$ | $0.10^{+0.03}_{-0.02}$ |
|  |  | 1400 | $(1.4^{+1.0}_{-0.5}) \times 10^{-12}$ | $0.011^{+0.007}_{-0.005}$ | $0.82 \pm 0.07$ |

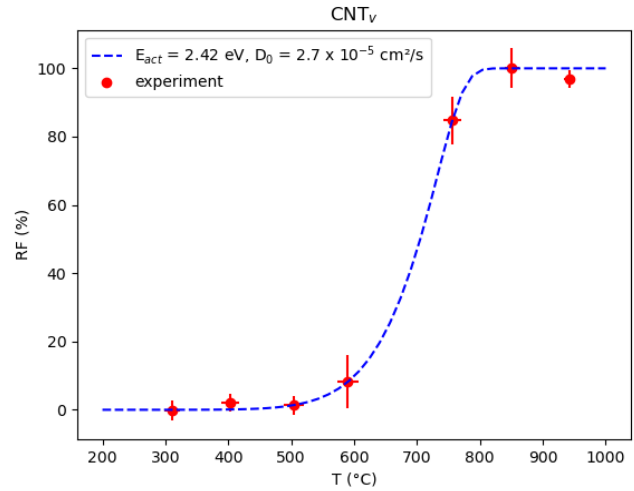

Figure 7: Comparison between the values of the experimental (red dots) and calculated (blue dotted lines) released fractions for the $CNT_v$ catcher. The calculated values are obtained with $D_0$ and $E_{act}$ given by the cylinder analysis.

The comparison between figures 6 and 7 shows that, for the $CNT_v$ catcher, the experimental RF value are reproduced as well by the sheet model as by the cylinder model. The comparison between the activation energies and $D_0$ obtained using either the catcher-sized sheet model or the carbon nanotube-sized cylinder model shows that the activation energies are close (2.68 and 2.3 eV or 2.42 eV) as expected as the carbon bonds are identical, and that $\sqrt{D_0}$ is reduced by a factor of $8.3 \times 10^4$ or $1.5 \times 10^4$ depending on the implantation depth assumed (33.8 or 90 μm respectively). These factors are close to the value expected ($1.8 \times 10^4$) considering the thickness of matter (or matter + free space) traversed, 90 μm in the case of the sheet and 5 nm in the case of the cylinder; the maximum remaining factor (4.6) results from the small difference between the activation energies found in the sheet and cylinder analyses and the strong dependence of $D_0$ on $E_{act}$ (see figure 5).

The values of $t_{50}$ and $\varepsilon_R(^{74}Rb)$ obtained at 1000 and 1400 °C using the sheet model in the extreme implantation conditions ($\delta$ = 33.8 and 90 μm) frame those calculated with the cylinder model (see tables 5 and 6). This shows that for the $CNT_v$ catcher, the release is mainly governed by diffusion in the CNT and that effusion plays a minor role.

It should be remembered that, as only 69 % of the Rb were stopped in the $CNT_v$, the effective release efficiency of this catcher is 56.6 % according to the cylinder model. If the CNTs had grown not on the quartz backing but on a releasing support, Rb diffusion could occur at the boundary between both media (CNTs and support). This diffusion rate depends on the diffusion coefficients in both media and on the size of the contact area between the CNTs and the support. The vertically aligned CNTs occupy only 8 % of the support surface, a contact area assumed lower than in the $CNT_h$-catcher case. The diffusion of Rb between the CNTs and the support would be minimised compared to what was observed with the $CNT_h$ catcher and the release efficiency of the support would be added to that of the CNTs in proportion to the atoms collected in the support. For example, if the $CNT_v$ had grown on a Papyex® support, the overall release efficiency for $^{74}$Rb at 1400 °C would be 71 %.

7. Conclusion

We studied the release profile of $^{81}$Rb out of four catchers (Papyex®, POCO, CNT$_v$ and CNT$_h$) with different microstructures in a temperature range from 250 °C to 950 °C. Two categories of release emerged, related to Papyex® and POCO materials on the one hand and CNT$_h$ and CNT$_v$ on the other.

The analysis of these release profiles was carried out in the framework of a diffusion model in a catcher-sized sheet. In order to carry out this analysis, the profile and the depth of the collected atoms were obtained using the LISE++ calculation code. In all the materials, the implantation was considered uniform.

In the case of CNT$_v$ and CNT$_h$ catchers, part of the atoms was collected in the support (quartz and Papyex® respectively). The release profile of the CNT$_v$ catchers showed that the atoms implanted in the quartz support were not released. In the case of the CNT$_h$ catchers, the profiles of the released fractions obtained for the Papyex® support on the one hand and for the CNT$_h$ on the other hand suggest an exchange of $^{81}$Rb between these two materials during the release, making the analysis of this catcher hazardous as the models used are diffusion in a sheet with release from one or two faces. Neither of them can describe an exchange between two materials.

For the other three catchers (Papyex®, POCO and CNT$_v$) the results of the analysis were used to extrapolate the release properties of $^{74}$Rb at 1400 °C (the temperature aimed for the TULIP TISS). In the case of graphite catchers, the implantation occurred in the first micrometers (8.9 μm for Papyex® and 5.6 μm for POCO). The release efficiencies at 1400°C for $^{74}$Rb were estimated to be 47 % and 48 % for the Papyex® and POCO catchers respectively. Despite the differences in microstructure (SSA, porosity, and thus density) between Papyex® and POCO, which necessarily lead to different contributions of effusion and diffusion in these two catchers, release efficiencies are similar showing a compensation between both release processes. The best release efficiency was obtained for the CNT$_v$ catcher (estimated at 82 % for $^{74}$Rb at 1400 °C). This catcher has a very high porosity ($\geq 92$ %). The diffusion plays a dominant role in the release process as shown by the double analysis performed in the framework of the sheet and cylinder models. In conclusion, carbon nanotube catcher (CNT$_v$) has very promising release properties due to its architecture. Once out of the tube by diffusion, the effusion time up to be out of the catcher can be neglected because the carbon nanotubes are aligned toward the outer surface of the catcher and the free space between the tubes is important, thus the tubes form no closed porosity. Therefore, an ideal catcher could consist of carbon nanotubes having a small diameter to minimize the diffusion and long enough and tilted to stop the nuclei.

Acknowledgements
This work is part of the TULIP project (Projet-ANR-18-CE31-0023) supported by the French National Agency of Research (ANR). The authors would like to thank the ALTO team and Sébastien Wurth (SPR team leader) for help in the implementation of the experimental devices and for the beam preparation.